\documentclass[twoside,12pt]{article}
\usepackage{epsfig}
\usepackage{graphics}

\newcommand{\be}{\begin{equation}}
\newcommand{\ee}{\end{equation}}
\newcommand{\bea}{\begin{eqnarray}}
\newcommand{\eea}{\end{eqnarray}}

\begin{document}

\title{  Improved Description of One- and Two-Hole Excitations \\ after Electron Capture in 163 Holmium\\
and the Determination of the Neutrino Mass. 
} 
\author{Amand Faessler $^{1}$, F. \v{S}imkovic$^2$, \\
\\
$^{1}$ Institute of Theoretical Physics, University of Tuebingen, Germany,\\
$^2$ INR, 141980 Dubna, Moscow Region, Russia and \\
Comenius University, Physics Dept., \ $SK-842 15$ \  Bratislava, 
Slovakia.}
\maketitle             

\begin{abstract} 
The atomic  pair $^{163}Ho$ and  $^{163}Dy$ seems due to the small Q value 
of about 2.5 keV the best case to determine the neutrino mass by electron capture. 
The bolometer spectrum measures the full deexcitation energy 
of Dysprosium (by X rays and Auger electrons plus the recoil of Holmium, 
which can be neglected).  The spectrum has an upper 
limit given by the Q value minus the neutrino mass. 
Till now this spectrum has been calculated allowing in Dysprosium excitations 
with 3s1/2, 3p1/2, 4s1/2, 4p1/2, 5s1/2, 5p1/2 (and 6s1/2) holes only. Robertson [R. G. H. Robertson, arXiv: 1411.2906v1] calculated 
recently also the spectrum with two electron hole excitations in Dy. 
He took the probability for the excitation 
for the second electron hole from work of Carlson and Nestor [T. A. Carlson, C. W. Nestor, T. C. Tucker, and F. B. Malik, Phys. Rev. 169, 27 (1968); T. A. Carlson and C. W. Nestor, Phys. Rev. A8, 2887 (1973)]  for Z=54 Xenon. 
The neutrino mass must finally be obtained by a simultaneous fit together with the Q value 
and the properties of the relevant resonances for the upper end of the spectrum. 
Under the assumption only one  
resonance (independent of its nature: one-, two-, multi-hole or of other origin,) near the Q value determines the upper 
end of the spectrum, and the profile of this leading state  
is Lorentzian, one has to fit simultaneously four parameters 
(neutrino mass, strength, distance of the leading resonance to the Q value and its width). 
If more than one resonance are of comparable importance for the upper end of the spectrum, 
it might be difficult or even impossible to
extract the neutrino mass reliably.  
Compared to the work of Robertson this work includes the following improvements:
(1) The two hole probabilities are calculated 
in the Dirac-Hartree-Fock (DHF) approach in Holmium 
and Dysprosium and not in Xenon. 
(2) In calculating the probability for the second electron hole in Dysprosium 
the n s1/2 or n p1/2 ($n\ \ge \  3$) one hole states 
are included selfconsistently in the DHF iteration. 
(3) Since Dysprosium has Z=66 electrons and Xenon only Z=54, one has at 
least 8 additional two hole states, which do not exist in Xenon and thus their probabilities 
have not been calculated by Carlson and Nestor and not been included by Robertson. 
They are included here.  
(4) For the probabilities of the one hole states, which determine the main 
structure of the spectrum, the overlap and exchange corrections are taken into account.  
(5) In solving the DHF electron wave functions the finite size of the nuclear 
charge distribution is included. 
(6) The nuclear matrix elements for electron capture integrates the 
charge of the captured electron over the nucleus 
with the weight $\psi(r)_{e, n,\ell, j} ^2\cdot r^2$.
For the capture probability thus the value $\psi_{e,n, \ell,j} ^2(R) \cdot R^2$ is taken 
at the nuclear radius 
and not the value $\psi_{e,n,\ell,j}^2(r = 0.0)\  at\  r\  =\  0.0$, 
which has the weight $r^2$ zero. 
(7) The formulas are derived in second quantization including automatically 
the antisymmetrization.
\end{abstract}

\vspace{1cm}

\section{Introduction}

The determination of the absolute value of the neutrino masses is one of the 
most important open problems in particle physics. Presently major efforts are underway 
to measure in the single beta decay, specifically in the Tritium decay, the 
electron antineutrino mass (KATRIN)  \cite{Drexlin}. 
The neutrinoless double beta decay can distinguish between
Dirac and Majorana neutrinos and is in principle able to measure the effective Majorana 
neutrino mass \cite{Fae2}. Electron capture measures the 
electron neutrino mass \cite{de,Fae3, Blaum, Alpert}. The sensitivity to the electron neutrino mass 
is increased (as in the single beta decay) in electron capture by a smaller Q value. 
Perhaps the best system for the determination of the absolute scale of the electron neutrino mass 
(and with the help of neutrino oscillations also of the muon and tauon neutrinos) by electron 
capture is the system $^{163}_{67}Ho$ and $^{163}_{66}Dy$. 

The electron neutrino mass can in principle be determined by the upper end of the 
deexcitation (bolometer) spectrum of $^{163} Dy$ after electron capture in $^{163}Ho$. All 
deexcitation spectra (X rays, Auger electron and the recoil of Holmium) 
end at the Q value  (for $^{163}Ho$ around 2.5 keV, 
see section 2 for the present Q value situation) 
minus the neutrino mass. {

Recently Robertson \cite{Robertson} included in the atomic excitations of 
Dysprosium also a second hole. The deexcitation bolometer spectrum 
including also the two-hole states adds to the 
leading contributions of the one-hole spectrum 
a "fine structure".   
For the probability of the second hole in Z=66 Dy Robertson  used results of 
Carlson and Nestor \cite{Carlson1, Carlson2} determined in Z=54 Xenon. 

The neutrino mass must be obtained 
by a simultaneous fit of the Q value and the neutrino mass to the upper end of the spectrum 
including also properties of one-hole, two-hole and other excitations close to Q. 
This complicates the neutrino mass determination. In case, that several resonances determine 
the upper end of the spectrum this can make the 
determination of the neutrino mass even impossible. If only one resonance determines the upper 
end of the spectrum and the line profile is Lorentzian, one has to fit four  parameters: 
neutrino mass, the distance of the leading resonance to the Q value, the strength, and the width. Before one fits the the theoretical spectrum to the upper end of the measurement, one has to fold the theory with the detector response including the finite resolution.
\\

The improvements compared to the work of Robertson \cite{Robertson}, Carlson and Nestor  
\cite{Carlson1, Carlson2}  are:  
 
\begin{itemize}
\item  The two hole probabilities are calculated here with a fully relativistic  
Dirac-Hartree-Fock (DHF) code of Grant \cite{Grant}, Desclaux \cite{Desclaux} and Ankudinov et al. 
\cite{Aukudinov} with the full antisymmetrization in the atoms Holmium 
and Dysprosium involved in the electron capture and not in Xenon. 

\item In calculating the probability for the second electron hole in Dysprosium 
the n s1/2 or n p1/2 ($ n \ \ge \ 3$) one hole states 
are included selfconsistently in the DHF iteration.
Thus for each one hole state in Dysprosium a full selfconsistent DHF iteration  
with a hole in the different one hole states is performed for all electrons in the atom. 
The electron wave functions in Ho and Dy for the same quantum numbers 
 $|n, \ell,j>$ but with different holes change markedly. In a very rough approximation 
for a first orientation without the Pauli correction the probability for the second hole is: 

\be
P(two-hole) \approx [1.0 - <Ho, n, \ell,j|Dy,n,\ell,j>^{2(2j+1)}] 
\label{P2H}
\ee

The overlap of electron orbitals with the same quantum numbers in Ho and Dy  are typically 
around 0.999 ( see eq. (\ref{over1}) and eq. (\ref{over2})). Thus a relative change in the overlap
by one percent  can change the two hole probability by a factor 10. Eq. (\ref{P2H}) serves as 
lever to enlarge a small change or error in the overlap into a large change in the two-hole 
probability. Therefore  the electron wave functions for the 
overlaps should be calculated in Holmium and 
Dysprosium and in the determination 
of the Dy electron wave functions the one hole state must be selfconsistently included, 
if one wants to obtain reliable probabilities for the two-hole states.
  
\item Since in Z=66 Dysprosium one has more electrons than in Z=54 Xenon, one has at 
least 8 additional two hole states (3s1/2 4f5/2; 3s1/2 4f7/2; 3p1/2 4f5/2; 
3p1/2 4f7/2; 4s1/2 4f5/2; 4s1/2 4f7/2; 4p1/2 4f5/2 4p1/2 4f7/2),  
which do not exist in Xenon and thus their probabilities 
have not been calculated by Carlson and Nestor \cite{Carlson1, Carlson2} 
and have not been included by Robertson \cite{Robertson}. 
They are included here. The excitation energies of 
these states are according to tables \ref{TwoHoles} and  \ref{TwoHoles2} 
around 2050, 1845, 415 and 335 eV.                                                          
Two hole states involving the one hole orbital 6s1/2 (P1) 
are not included here, because already the 5s1/2 has only an excitation energy of 44.7 eV. 
The 6s1/2 excitation energy is expected around 5 eV, i. e. 100 times smaller 
than the excitations neglected by Robertson \cite{Robertson}. 
In addition the ionization energy 
of the 6s1/2 state in the atoms involved seems not to be available in the literature 
and these very lightly bound electrons are not well 
described by Slater determinants due to configuration mixing.

\item The electron wave functions in the parent 
and in the daughter atom get more and more similar with increasing charge number Z, 
since the relative change $\Delta Z/ Z$ is smaller. Thus the overlaps  
of the wave functions for Z in the parent and (Z-1) in the daughter increase 
closer to unity 
and the sudden approximation yields a smaller two hole excitation probability,
which is very roughly given by eq. (\ref{P2H}). 
The overlaps are tabulated in Faessler, Gastaldo and Simkovic \cite{Fae3}. They  
are close to unity. Thus the change of these overlaps 
from Xenon to Holmium-Dysprosium needs only 
to be one percent to yield a difference of a factor ten or more  for the 
 two-hole probabilities. For one hole in the Dysprosium state 3s1/2 or 4s1/2 : 

\begin{eqnarray} 
\mbox{hole in 3s1/2:   } <Ho,3s1/2|Dy,3s1/2>\  =\  0.999390; \nonumber  \\
 <Ho,4s1/2||Dy,4s1/2> \ = \ 0.999332; 
\label{over1}
\end{eqnarray}

\begin{eqnarray}
\mbox{hole in 4s1/2:   } <Ho,3s1/2|Dy,3s1/2> \ = \ 0.999377; \nonumber \\
<Ho,4s1/2|Dy,4s1/2> \ = \ 0.998870; 
\label{over2} 
\end{eqnarray}

(All numbers in this work are calculated in double precision.) 
  
\item For the probabilities of the one hole states 3s1/2 (M1), 3p1/2 (M2), 4s1/2 (N1), 4p1/2 (N2),
    5s1/2 (O1), 5p1/2 (O2), 6s1/2 (P1), which determine the main 
structure of the spectrum, the overlap and exchange corrections are included 
according to Faessler, Gastaldo and Simkovic \cite{Fae3}. 

\item In solving the DHF electron wave functions \cite{Fae2,Grant, Desclaux, Aukudinov} 
the nuclear 
charge distribution is included by the Fermi parametrization  determined 
by electron-nucleus scattering.
 
\item The nuclear matrix element for electron capture integrates the 
charge of the captured Ho electron over the nucleus with the weight $\psi^2(r)_{e, n,\ell, j}\cdot r^2$.
For the capture probability thus the value $\psi^2(R)_{e,n, \ell,j} \cdot R^2$ is taken 
 (R = nuclear radius) and not the value 
$\psi^2(r = 0.0)_{e, n,\ell, j}$ at r = 0.0, which has the $r^2$ weight zero. 

\item The formulas are derived in second quantization including automatically 
the full antisymmetrization. This formulation allows not only to describe two-hole states 
but also to extend the description for three holes and even more hole states. 
\end{itemize}

\vspace{1cm}

\section{Description of Electron Capture \\ and the Atomic Wave Functions}

The bolometer spectrum of the deexcitation of $^{163} Dy$  
after electron capture in $^{163} Ho$ can be expressed as \cite{de} and \cite{Fae3} 
assuming Lorentzian line profiles: 

\be
 \frac{d\Gamma}{dE_c} \propto \sum_{i = 1,...N_\nu}(Q - E_c)
\cdot U_{e,i}^2\cdot\sqrt{(Q-E_c)^2 -m_{\nu,i}^2} \\
*\sum_{f=f'} \lambda_{0}B_f \frac{\Gamma_{f'}}{2\pi} 
\frac{1}{(E_c - E_{f'})^2 +\Gamma_{f'}^2/4} \label{decay}
\ee
With $Q \  = \ 2.3 \ to \ 2.8 \ keV $ \cite{Blaum, 
Anderson, Gatti, Ra, Audi}, with a recommended  
value \cite{Wang}  $Q = (2.55 \pm 0.016)$  keV,  $U_{e,i}^2$ 
the probability for the admixture of different 
neutrino mass 
eigenstates $i\ =\ 1,..N_\nu$  into electron neutrinos 
and $E_c$ is the excitation energy of final Dysprosium;
$B_f$ are the overlap and exchange corrections; $\lambda_{0}$ contains 
the nuclear matrix element squared \cite{bam}; 
$ E_{f'}$ are the one- and two-hole excitation energies in Dysprosium; 
$ \Gamma_{f'}$ are the widths of the one- and two-hole states in Dysprosium \cite{Fae3}.

Here as in all other calculations for the deexcitation of Dy after electron 
capture a Lorentzian shape is assumed. This is probably a good description. 
Holmium is in the ECHo experiment built in a gold film positioned as an interstitial 
 or it occupies a position of the gold lattice. A Gaussian shape would be 
expected in a gas from Doppler broadening. Even collision and pressure broadening 
yield usually a Lorentzian profile. But since the shape of the resonance lines are 
important for the determination of the neutrino mass, the line shapes should 
be studied in the future more carefully.  

Results of the ECHo collaboration \cite{Ra} yield 
for electron capture in $^{163}Ho$ to $^{163}Dy$ a Q-value:

\be
Q(ECHo)\  =\  (2.80\pm 0.08)\  keV.
\label{QECHo}
\ee\ 
The highest two hole state in Dy has an energy of 2.474 keV (see table \ref{TwoHoles}) 
far below the Q value. So three-hole and multi-hole states (and perhaps also states from configuration mixing), 
which can be higher in energy, might be more dangerous for the determination of the neutrino mass.  
 
We assume, that the total atomic wave function 
can be described by a single Slater determinant. 
$ B_f $ takes into account the overlap and the exchange 
terms between the parent $ |G> $ and the daughter atom in 
the state $ |A'_f> $ with a hole in the electron 
state $|f'>$. We use the sudden approximation as Faessler et al. \cite{Fae1}. 
$ B_f$  is the overlap and exchange correction 
for the electron capture probability from the state f relative to the capture from $3s_{1/2}$ 
in Ho with one hole in f' in the Dy atom, given in eq. (\ref{Bff}) 
in the Vatai approximation \cite{Vatai1, Vatai2}. But the numerical value 
used here are calculated with the full overlap and 
exchange corrections of Faessler et al. \cite{Fae3}. 

\begin{eqnarray}
 B_f = \frac{|\psi_{f}(R) <A'_f|a_f|G>|^2}{|\psi_{3s1/2}(R)|^2} 
 = P_f \cdot \frac{|\psi_{f}(R)|^2}{|\psi_{3s1/2}(R)|^2}  
\label{Bff}
\end{eqnarray}

For two-hole final states one has to multiply eq.(\ref{Bff}) according to (\ref{Prob}) 
with the probability to form a second hole characterized by the quantum numbers "p' ". 
One has to replace $<A'_{f'}|a_i|G>$ by \\ $<A'_{f',p'; q'}|a_i|G>$
with the two electron holes f' and p' and the additional electron particle q' in Dysprosium
 above the Fermi surface F. 
The probability for the leading expression ( Wick \cite{Wick} contracted)  
to form one hole in Dy in f' = f = i is: 

\be
 P_f = |<A'_{f'}|a_{i = f }|G>|^2 \approx \prod_{k= 1, ...Z; \ne f} |<k'|k>|^2  \label{Bfff}
\ee

The corresponding probability  for two final hole states f' and p' and an additional
electron in q' summed 
over  all $q' \ >\  F$  of the unoccupied bound and the continuum states is: 

\be
 P_{p'/f'} = \sum_{q' > F} |<A'_{f', p'; q'}|a_i|G>|^2  \label{Bffff}
\ee
The antisymmetrized Slater determinants for the  wave functions of the initial Holmium in the
 ground state $ |G> $ and the excited one electron hole states $ |A'_f> $ in 
Dysprosium read in second quantization:

\be
  |G> =  a_1^{\dagger} a_2^{\dagger} a_3^{\dagger}... a_Z^{\dagger} |0> \label{G}
\ee

\be
  |A'_f> =  a'^{\dagger}_1 a'^{\dagger}_2... a'^{\dagger}_{f'-1}a'^{\dagger}_{f'+1}... 
	a'^{\dagger}_{Z} |0> \label{A}
\ee

The antisymmetrized two-hole state in Dy is:
\be
  |A'_{p',f'}> =  a'^{\dagger}_1 a'^{\dagger}_2... a'^{\dagger}_{f'-1}a'^{\dagger}_{f'+1}... 
	a'^{\dagger}_{p'-1}a'^{\dagger}_{p'+1}...a'^{\dagger}_{Z}a'^{\dagger}_{q' > F} |0> \label{A2}
\ee

The primes with the dagger indicate the single electron 
spinor creation operators for the daughter nucleus 
(Dysprosium) with one electron hole in the single particle state $|f'> $ in eq. (\ref{A}) and
two holes $|f'>$ and $|p'>$ in eq. (\ref{A2}). 
The following expressions have to be calculated 
with the help of Wick's theorem \cite{Wick}. One considers the Wick contractions as an expansion 
with  the "non-diagonal" overlaps as small parameters. The leading term is 
the expression without "non-diagonal" overlaps.  

\be
  P_f = |<A'_f|a_i|G>|^2 =  |<0| a'_Z a'_{Z-1}...
	a'_{f+1}a'_{f-1}...a'_1 \cdot a_f \cdot 
	a_1^{\dagger} a_2^{\dagger} a_3^{\dagger}... a_Z^{\dagger} |0>|^2 \label{Wick}
\ee 

\be
  P_{p/f} = |<A'_{p',f'}|a_f|G>|^2 =  \sum_{q' > F}|<0| a'_{q'}a'_Z ...a'_{p'+1}a'_{p'-1}...
	a'_{f+1}a'_{f-1}...a'_1 \cdot a_f \cdot 
	a_1^{\dagger} a_2^{\dagger} a_3^{\dagger}... a_Z^{\dagger} |0>|^2 \label{Wick2}
\ee 

The leading expression (without the sub-leading  exchange terms, which contain at least
 one "non-diagonal" overlap in the amplitude) is obtained in eq. (\ref{Wick})  
for electron capture in Holmium from the state i, 
if the captured electron i in Ho  has the same quantum numbers 
$n,\  \ell \ and \ j $ as the final hole state f' in Dy, i.e. for the quantum numbers i = f = f'. 
The probability for two hole states eq. (  \ref{Wick2}) is next to the leading order, since it contains 
always at least one "non-diagonal" overlap $<q'|p>$ for the amplitude and the square for the probability. 
If the sequence of f' and p' is reversed in Dysprosium 
the Fermion creation operators produce automatically a "-" sign 
for the amplitude, which does not matter for the probability.  

\section{Derivation of the Probabilities}

It has been already stressed, that the leading contribution 
for the one-hole state in eq. (\ref{Wick}) and 
for the two hole state in eq. (\ref{Wick2}) is obtained, 
if the orbital quantum numbers for the captured electron in Holmium 
 $|i>\ =\ |n,\ell ,j, m>$ are the same as the hole quantum numbers in Dy 
$|f'> \ = \ |n',\ell',j', m'>$, thus $i \ =
 \ f' \ = \ f$ and $n \ =\ n', \ \ell \ = \ \ell',\  j \ = \ j'$ and $m\ = m'$. 

The leading expression for the  probability to excite the hole state in Dy 
corresponding to the quantum numbers of the state of the captured electron in Holmium is:

\begin{eqnarray}
 P_f  = |<A'_f|a_f|G>|^2 = \prod_{k = k' <F, \ne f} <k'|k>|^2 = 
\prod_{(n, \ell, j)(Ho = Dy)<F_{Dy}} |<(nlj)_{Dy}|(n, \ell, j)_{Ho}>|^{2 \cdot N_{n, \ell, j}}                    
 \label{Ff}
\end{eqnarray}

The definition of $N_{n, \ell, j}$ is given in eq. (\ref{P4}). $(n, \ell, j)_{Ho}$ and 
$(n, \ell, j)_{Dy}$ indicate electrons in the Ho and in the Dy atom with the same quantum numbers.
Such states have a large overlap only slightly below unity. 

For the following one needs some elementary laws of probability calculus:

\bea
P(A \ and \ B) = P(A)\cdot P(B) \hspace{5cm} \nonumber \\
P(A\  or\  B)  = P(A) + P(B), \mbox{ if A and B exclude each other.} \hspace{2cm}
\label{Prob}
\eea

If one wants to include the leading contribution for two-holes in the  final 
Dy atom,  
one must multiply expression $P_f$ in eq. (\ref{Ff}) 
with the probability $P_{p/f}$ to form an additional  electron particle 
$q' \ > \  F;( \ F\ =\ $Fermi Surface) and electron hole state $ p' \ < \ F$.  
The excited electron q' can be in an unoccupied bound state or in the continuum of the Dy atom. 

\begin{eqnarray}
 P_{p/f}(q'>F) = |<0|a'_{q'}a'_Z...a'_{p'+1}a'_{p'-1}...
a'_{f'+1}a'_{f'-1}...a'_{1'}\cdot a_f \cdot a^+_1...a^+_Z|0>|^2 =  \nonumber \\
|<A'_{p',f'<F; q'>F}(2\ holes)|a_f|G>|^2 
\approx   |<q'_{>F}|p_{<F}>  \cdot \prod_{k=1..Z \neq f,p}<k'|k> |^2  
\label{con}
\end{eqnarray}

q' is an empty electron orbit in Dy, into which the electron p is scattered,  
and p the occupied state in Ho, from which this electron is removed. 
Here again k and k' and also f and f' and p and p' stand 
for the same electron quantum numbers $ n, \ \ell,\ j $ in the parent k, f, p and the 
daughter atom k', f', p'. The product over k runs 
over occupied states  $k'\ = \ k =(n_k,\ \ell_k,\ j_k,\ m_k)$ in Ho and  Dy with the exemption of 
$f\ = (\ n_f,\  \ell_f, \ j_f, \ m_f)$.  q' an  
empty state in Dy can be bound or in the continuum. If q' is in the  continuum,   
one speaks of `shake off'. 
Since now a "non-diagonal" overlap is involved in eq. (\ref{con}) 
with $<q'_{Dy}|p_{Ho}> $ already in the amplitude 
and this expression must be squared for the probability, the two hole 
contributions are reduced by a "non-diagonal" overlap squared. 
If one exchanges the states f' and p', one obtains an additional "-" sign . But since one 
has to square the expression, a phase is irrelevant.

To evaluate the probability for an additional electron 
particle-hole state  (\ref{con}) 
one sums incoherently over all unoccupied states q'.  This assumes, that 
the different two-hole excitations do not influence each other. 

\bea
P_{f', p'} =  \sum_{q' > F} |<p_{<F, Ho}|q'_{>F, Dy}><q'_{>F, Dy}|p_{<F, Ho}>|
\cdot \prod_{k=k'<F_{Dy} \neq f,p}|<k'_{Dy}|k_{Ho}>|^2 
\label{F22}
\eea

Here as stressed above the sum over q' runs only over the unoccupied bound and continuum 
states in Dy. 
One can now use the completeness relation to shift the sum 
over q' to states away from the continuum to states, 
which one can calculate easier. One divides the completeness
relation into two pieces: up to the last occupied state below the Fermi Surface  F  
and all states above the last occupied state  
including also the continuum. 

\be
 1 = \sum_{q' < F} <p|q'><q'|p>  + \sum_{q'>F} <p|q'><q'|p>
\label{comp}
\ee

The sum in eq. (\ref{F22}) is the last part of 
eq. (\ref{comp}) and one can transcribe (\ref{F22}) into:

\bea
P_{p/f} = \left( 1 - \sum_{q'<F} <p_{Ho}|q'_{Dy}><q'_{Dy}|p_{Ho}>\right)
\prod_{k=k' < F_{Dy}; \ne p, f} 
<k'_{Dy}|k_{Ho}>  
\label{P3}
\eea

In the literature one uses often the Vatai 
approximation \cite{Vatai1, Vatai2}: Exchange corrections 
have been already neglected in the previous expressions. In addition one assumes, 
that the overlaps of electron wave functions in the parent and the daughter atom 
with the same quantum numbers (given in the product term in eq. (\ref{P3})) 
can be approximated by unity. Typically the overlaps 
have values \cite{Fae3} 
$ <k'_{Dy}|k_{Ho}> \approx 0.999$ and thus for Dy minus two 
holes $0.999^{64} \ \approx \ 0.94$. In the 
Vatai approximation \cite{Vatai2} one replaces this value by 1.0.

\bea
P_{p/f} = \left( 1 - \sum_{q'<F} <p_{Ho}|q'_{Dy}><q'_{Dy}|p_{Ho}>\right) 
=   \hspace{4cm} \nonumber \\
\left( 1 -<p_{Ho}|p'_{Dy}><p'_{Dy}|p_{Ho}> - 
\sum_{q'<F, \ne p'} <p_{Ho}|q'_{Dy}><q'_{Dy}|p_{Ho}>\right) 
\label{P2} 
\eea

 The physics of the two terms subtracted from 1 in eq. (\ref{P2}) is: 
The first subtracted term gives the probability, that the state p' in Dy is occupied.
The second terms take into account the Pauli principle and prevent, that electrons 
can be moved into occupied states in Dy. The single electron states like 
$|p> = |n, \ell, j, m> $ include also 
the angular projection quantum number m. This projection is for 
the description of the data irrelevant. The first subtracted term in eq. (\ref{P2}) 
gives the probability, that a specific magnetic substate m' is occupied in p'. The probability, 
that all magnetic substates of p' are occupied is an "and" situation (\ref{Prob}) 
and the probabilities for the substates have to be multiplied and one obtains the Nth. 
power of the single electron probability with $N_{p'} = N_{n,\ell,j; p'} = (2j + 1)_{p'}$: \ \  
$ |<(n,\ell, j)_{p,Ho}|(n,\ell,j)_{p'.Dy}>|^{2 \cdot N_{n,\ell,j; p'}}$.

\bea
P_{p/f} = ( 1 - |<(n,\ell,j)_{p, Ho}|(n,\ell,j)_{p',Dy}>|^{2  \cdot N_{p'}} - 
\nonumber \\
\sum_{(n, \ell, \ j)_{q',Dy; \ne p'} <F} \frac{N_{n,\ell,j}\cdot N_{n',\ell,j}}
{2j + 1}
 |<(n,\ell,j)_{p, Ho}|(n,\ell,j)_{q',Dy}>|^2  ) 
\label{P1}
\eea

\be
\mbox{for n and n' with:  } \hspace{1cm} N_{n,\ell,j} = 2j +1;   
\ee

But for the primary hole state f and the 4f7/2 state one has special factors: 

\be     
N_{(n,\ell,j)_f} = 2j_
f \hspace{0.5cm} and \hspace{0.5cm} N_{4f7/2} = 5. 
\label{P4}  
\ee

Here $|f>$  
is the orbital in Ho from which the electron is originally captured. 
The electron p' is moved to $4f7/2$ with now 5 electrons in this orbit to guarantee the correct number of electrons.    
$N_{n,\ell,j}/(2j +1)$ is the averaged probability to find an electron 
in the $|p,n,\ell,j,m>$ orbital and $N_{n',\ell,j}$ the number of electron in the 
$|q', n',\ell,j>$ state. 
For the probability of the second hole we used the 
Vatai \cite{Vatai1,Vatai2} approximation with  the overlaps of 
corresponding electron wave functions with the same quantum numbers in Ho and Dy equal to unity 
and neglected the exchange corrections. For the "diagonal" overlaps of 
the order of $ \approx 0.999$ this is a good approximation. 

\begin{table}
\caption{ One- and two-hole states in $^{163}Dy$ with quantum numbers $n, \ell, j$. 
$E_C$ is the excitation energy of the one- or the two-hole state. 
We adopt here for a better comparison the values for $E_c$ and the width $ \Gamma$  
used by Robertson in his arXiv publication \cite{Robertson}
taken from  \cite{Weast}, although there seem 
to be in some cases better values in the literature 
\cite{Ra}, \cite{Deslattes}, \cite{Thompson}, \cite{Campbell} and \cite{Cohen}. The 
last two columns give the relative probabilities of the one and the two hole 
states in relation to $3s1/2$ in $\%$ for $^{163}Dy$ of the present work (P-Fae) 
and the publication of Robertson (P-Rob) \cite{Robertson}. One finds surprisingly 
large differences for the two-hole probabilities to Robertson \cite{Robertson}. 
To ensure, that the present 
values are correct, three two-hole probabilities with about 
the largest differences to Robertson have also been calculated by hand. The two-hole 
probabilities are extremely sensitive to the overlap between the relativistic electron orbitals 
in Ho and Dy with the same quantum numbers $|n, \ell, j>$. In a very rough 
approximation eq. (\ref{P2H}) provides a lever to enlarge a small change for the overlap 
into a large change in the two-hole probability. The difference between the Robertson results 
and the present approach is largest for the outermost electron orbitals for the second hole. 
One expects for the slightly bound states  5s1/2, 5p1/2 and 5p3/2 for the overlaps in 
Xe and in Ho the largest differences, since these states are much weaker bound 
in Xe than in Ho. See table \ref{Lever}. }
\label{TwoHoles}
\begin{center}

\begin{tabular}{|c |r|r|r|r|r|} \hline
$1.\ hole$ & $2.\  hole$ & $E_c [eV]$  & $\Gamma[eV] $& $P-Fae [\%]$ 
& $P-Rob [\%] $\\ \hline \hline
 3s1/2 &  ----- & 2041.8  & 13.2 &   100   & 100   \\ \hline
 3s1/2 &  4s1/2 & 2474.2  & 13.2 &   0.167 & 0.075 \\ \hline
 3s1/2 & 4p1/2  & 2385.3  & 13.2 &   0.103 & 0.11   \\ \hline
 3s1/2 & 4p3/2  & 2350.0  & 13.2 &   0.163 & 0.25   \\ \hline
 3s1/2 & 4d3/2  & 2201.8  & 13.2 &   0.930 & 1.05  \\ \hline
 3s1/2 & 4d5/2  & 2201.8  & 13.2 &   0.126 & 1.62  \\ \hline
 3s1/2 & 4f5/2  & 2050.4  & 13.2 &   0.165 & 0.0   \\ \hline
 3s1/2 & 4f7/2  & 2047.0  & 13.2 &   0.182 & 0.0   \\ \hline
 3s1/2 & 5s1/2  & 2091.1  & 13.2 &   0.132 & 1.36  \\ \hline
 3s1/2 & 5p1/2  & 2072.6  & 13.2 &   0.128 & 3.12  \\ \hline
 3s1/2 & 5p3/2  & 2065.9  & 13.2 &   0.185 & 7.31  \\ \hline \hline 
 3p1/2 & -----  & 1836.8  &  6   &   5.080 & 5.26  \\ \hline
 3p1/2 & 4s1/2  & 2269.2  &  6   &   0.005 & 0.004  \\ \hline
 3p1/2 & 4p1/2  & 2180.3  &  6   &   0.009 & 0.006  \\ \hline
 3p1/2 & 4p3/2  & 2145.0  &  6   &   0.007 & 0.014 \\ \hline
 3p1/2 & 4d3/2  & 1996.8  &  6   &   0.005 & 0.057  \\ \hline
 3p1/2 & 4d5/2  & 1996.8  &  6   &   0.005 & 0.087 \\ \hline
 3p1/2 & 4f5/2  & 1845.4  &  6   &   0.005 & 0.00  \\ \hline
 3p1/2 & 4f7/2  & 1842.0  &  6   &   0.012 & 0.00  \\ \hline
 3p1/2 & 5s1/2  & 1886.1  &  6   &   0.005 & 0.072  \\ \hline
 3p1/2 & 5p1/2  & 1887.6  &  6   &   0.008 & 0.165  \\ \hline
 3p1/2 & 5p3/2  & 1860.9  &  6   &   0.008 & 0.386  \\ \hline \hline
\end{tabular}
\end{center}
\end{table}

\vspace{1cm}

\begin{table}
\caption{ Continuation of the table \ref{TwoHoles} 
with one and two hole states in $^{163}Dy$.}
\label{TwoHoles2}
\begin{center}

\begin{tabular}{|c |r|r|r|r|r|} \hline
$1.\ hole$ & $2.\  hole$ & $E_c [eV]$  & $\Gamma[eV] $& $P-Fae [\%]$ 
& $P-Rob [\%] $\\ \hline \hline
 4s1/2 & -----  &  409.0  &  5.4 &  24.40  &23.29  \\ \hline
 4s1/2 & 4s1/2  &  841.4  &  5.4 &   0.021 & 0.001  \\ \hline
 4s1/2 & 4p1/2  &  752.5  &  5.4 &   0.052 & 0.004  \\ \hline
 4s1/2 & 4p3/2  &  717.2  &  5.4 &   0.091 & 0.01  \\ \hline 
 4s1/2 & 4d3/2  &  569.0  &  5.4 &   0.088 & 0.077  \\ \hline
 4s1/2 & 4d5/2  &  569.0  &  5.4 &   0.125 & 0.123  \\ \hline
 4s1/2 & 4f5/2  &  417.6  &  5.4 &   0.027 & 0.0  \\ \hline
 4s1/2 & 4f7/2  &  414.2  &  5.4 &   0.023 & 0.0  \\ \hline
 4s1/2 & 5s1/2  &  458.3  &  5.4 &   0.066 & 0.254 \\ \hline
 4s1/2 & 5p1/2  &  439.8  &  5.4 &   0.039 & 0.629 \\ \hline
 4s1/2 & 5p3/2  &  433.1  &  5.4 &   0.058 & 1.502 \\ \hline \hline
 4p1/2 & -----  &  328.3  &  5.3 &   1.220 & 1.19 \\ \hline
 4p1/2 & 4p1/2  &  671.8  &  5.3 &   0.001 & 0.0001 \\ \hline
 4p1/2 & 4p3/2  &  636.5  &  5.3 &   0.004 & 0.0005 \\ \hline
 4p1/2 & 4d3/2  &  488.3  &  5.3 &   0.005 & 0.004 \\ \hline
 4p1/2 & 4d5/2  &  488.3  &  5.3 &   0.006 & 0.006 \\ \hline
 4p1/2 & 4f5/2  &  336.9  &  5.3 &   0.002 & 0.0 \\ \hline
 4p1/2 & 4f7/2  &  328.3  &  5.3 &   0.001 & 0.0 \\ \hline
 4p1/2 & 5s1/2  &  377.6  &  5.3 &   0.002 & 0.013 \\ \hline
 4p1/2 & 5p1/2  &  359.1  &  5.3 &   0.004 & 0.031 \\ \hline
 4p1/2 & 5p3/2  &  352.4  &  5.3 &   0.002 & 0.076 \\ \hline \hline
 5s1/2 & 5s1/2  &   44.7  &  3   &   3.200 & 3.45 \\ \hline \hline
 5p1/2 & 5p1/2  &   21.1  &  3   &   0.157 & 0.15 \\ \hline \hline
\end{tabular}
\end{center}
\end{table}
\begin{table}
\caption{ Ionization energies and two-hole probabilities 
$P[\%]$ (\ref{Bff}) relative to the 3s1/2 state comparing Xe with Ho. 
A small reduction  of the overlap, which is typically for well bound electron 
orbitals $ <Z,\ n,\  \ell,\ j\ |\ Z-1,\ n, \ \ell, \ j> \ = \ 0.999 $, 
by an assumed value of 2 \%
produces by the lever of eq. (\ref{P2H}) a large increase of the second hole probability.  
The reduction of the overlap for going from an atom with Z to Z-1 is largest 
for very lightly bound electrons. The $ 5s1/2, \ 5p1/2 \  and \  5p3/2$ are 
in Xenon with Z = 54 only with half the energy E bound as in Holmium with Z = 67. 
The binding energies of the Holmium orbitals are taken from Robertson \cite{Robertson}. 
The values in the brackets are from the 
literature \cite{Weast, Deslattes, Thompson, Campbell, Cohen}. 
A decrease of the overlaps in Xenon by assumed $2 \ \%  \ to \ 0.979$ relative to Holmium 
increases the probability for the 
second hole P(2-hole) by a factor 20. This factor increases the 2-hole probabilities 
in these states to roughly the values of Robertson \cite{Robertson} 
(see table \ref{TwoHoles} and \ref{TwoHoles2}), who used the 
results of Carlson and Nestor \cite{Carlson2} calculated for Xenon. It should be stressed, 
that the two-hole probabilities are in this work not calculated by the very rough 
eq. (\ref{P2H}) but by the more accurate expression (\ref{P1}).
\label{Lever}}
\begin{center}
\begin{tabular}{|c |c|c|} \hline
 -       &  $ E\ [eV]\ Xenon$  & $E\ [eV]\ Holmium$   \\ \hline \hline
$5s_{1/2}$    & 23.3  &    49.9     \\ \hline
$5p_{1/2}$    & 13.4  & 26.3 (30.8) \\ \hline
$5p_{3/2}$    & 12.1  & 26.3 (24.1) \\ \hline  \hline
$Overlap $    & $for \ (-2\ \%)$ 0.979; P(2-hole) & $for$ 0.999; P(2-hole) \\ \hline \hline
$j\ = \ 1/2$  &  8.1 \%  & 0.4 \%   \\ \hline 
$j\ = \ 3/2$  & 15.6 \%  & 0.8 \%   \\ \hline \hline 
\end{tabular}
\end{center}
\end{table}

\begin{itemize}

\item In the formation of the first hole state in orbital f in Ho and f'
in Dy the electron can 
be captured from each projection quantum number "m". 
One has an "or" situation. According to eq. (\ref{Prob}) the probabilities add up. 
But since one has capture only from $s_{1/2}$ and $p_{1/2}$ states, this yields a common 
factor two in the amplitude and thus 
is not relevant for the relative probabilities.

\item The probability amplitude for the second hole is calculated in the sudden approximation 
\cite{Fae1} by the overlap of the ground state of Ho with the captured electron f removed 
with Dy with a hole in f' and the particle-hole excitation 
 ($q'^1 \cdot p'^{-1}$). The overlap squared 
$ |<(n,\ell, j)_{p,Ho}|(n,\ell,j)_{p'.Dy}>|^2$ gives the probability, 
that a specific magnetic substate $m_{p'}$ is occupied. To obtain the probability, 
that all magnetic substates of p' are occupied, one has the "and" situation (\ref{Prob}), which 
requires to multiply all these probabilities with each other, 
which yields the power of $N_{(n,\ell,j) p'}$. 

\item The second subtracted term in eq. (\ref{P1}) takes care of the 
Pauli principle of the occupied states apart of p'. 
One has an $"or"$ situation. Thus one obtains a factor $N_{n,\ell,j}$ according to eqs.
(\ref{Prob}) and (\ref{P4}). 

\item The one hole f' and the particle-hole excitation $(q'^1 \cdot p'^{-1})$   
are in  an "and"  situation. Thus the probabilities have to be multiplied  
$P(f'^{-1})\cdot P(q'^1,p'^{-1})$ 
according to ( \ref{Prob}). One multiplies  the probabilities (\ref{Ff}) with (\ref{con}). 
These probabilities relative to 3s1/2 (\ref{Bff}) are listed in tables \ref{TwoHoles} and \ref{TwoHoles2}. 
\end{itemize}  

\begin{table}
\caption{ The electron binding energies and widths of hole states in $^{163}Ho$ 
from the literature 
\cite{Weast,Deslattes,Thompson,Campbell,Cohen} and 
the recent ECHo data \cite{Ra2,Lo2}. Electrons below $3s_{1/2}$ can not be captured in 
$^{163}Ho$. Due to the Q-value of about 2.8 keV they are energetically forbidden. 
\label{Bind}}
\begin{center}
\begin{tabular}{|c |c|r|r|r|r|} \hline
 -       &  $ n,\ell,j$  & $E_{lit}[keV] $   &   $ E_{ECHo}[keV]$ 
& $\Gamma_{lit}[eV]$ &$\Gamma_{ECHo}[eV]$ \\ \hline \hline
$M1$       &$3s{1/2}$    & 2.047  & 2.040   & 13.2 &  13.7 \\ \hline
$M2$      &$3p_{1/2}$    & 1.836  & 1.836   & 6.0  &  7.2  \\ \hline
$N1$      &$4s_{1/2}$    & 0.420  & 0.411   & 5.4  &  5.3  \\ \hline
$N2$      &$4p_{1/2}$    & 0.340  & 0.333   & 5.3  &  8.0  \\ \hline
$O1$      &$5s_{1/2}$    & 0.050  & 0.048   & 5.0  &  4.3  \\ \hline \hline
\end{tabular}
\end{center}
\end{table}

\vspace{1.0cm}

\begin{figure}[tp]
\begin{center}
\begin{minipage}[tl]{18 cm}
\epsfig{file=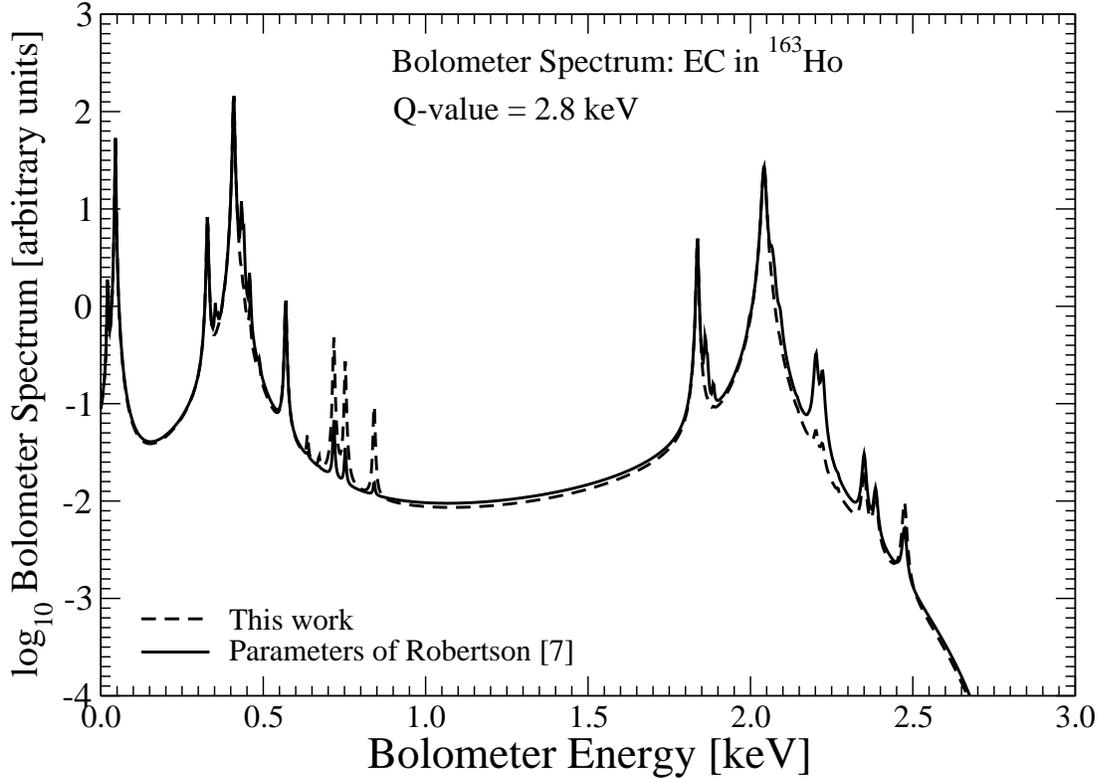,angle= -90,scale=0.6}
\end{minipage}
\begin{minipage}[t]{16.5 cm}
\caption{$Logarithmic_{10}$ (with basis 10) bolometer Spectrum (\ref{decay}) and (\ref{P1}) for the one-  
and two-hole probabilities calculated in this work (dashed line) and  
with the parameters of Robertson (solid line) \cite{Robertson} with the assumed Q-value = 2.8 keV for
the bolometer energy between 0.0 and 2.8 keV. Apart of the excitation of one-hole states in
$^{163}Dy$ after electron capture in $^{163}Ho$   
also the excitations of two-hole states in $^{163}Dy$ are included. The $logarithmic_{10}$  
plot stresses the effect of the the two hole states. In a linear plot 
(see figure \ref{Lin}) the two hole states are  
hardly to be seen. The values at the ordinate have to be read as $10^{ordinate}$. 
So "-2" is $10^{-2}$.    
\label{FaeRob}}
\end{minipage} 
\end{center}
\end{figure}
 
\begin{figure}[tp]
\begin{center}
\begin{minipage}[tl]{18 cm}
\epsfig{file=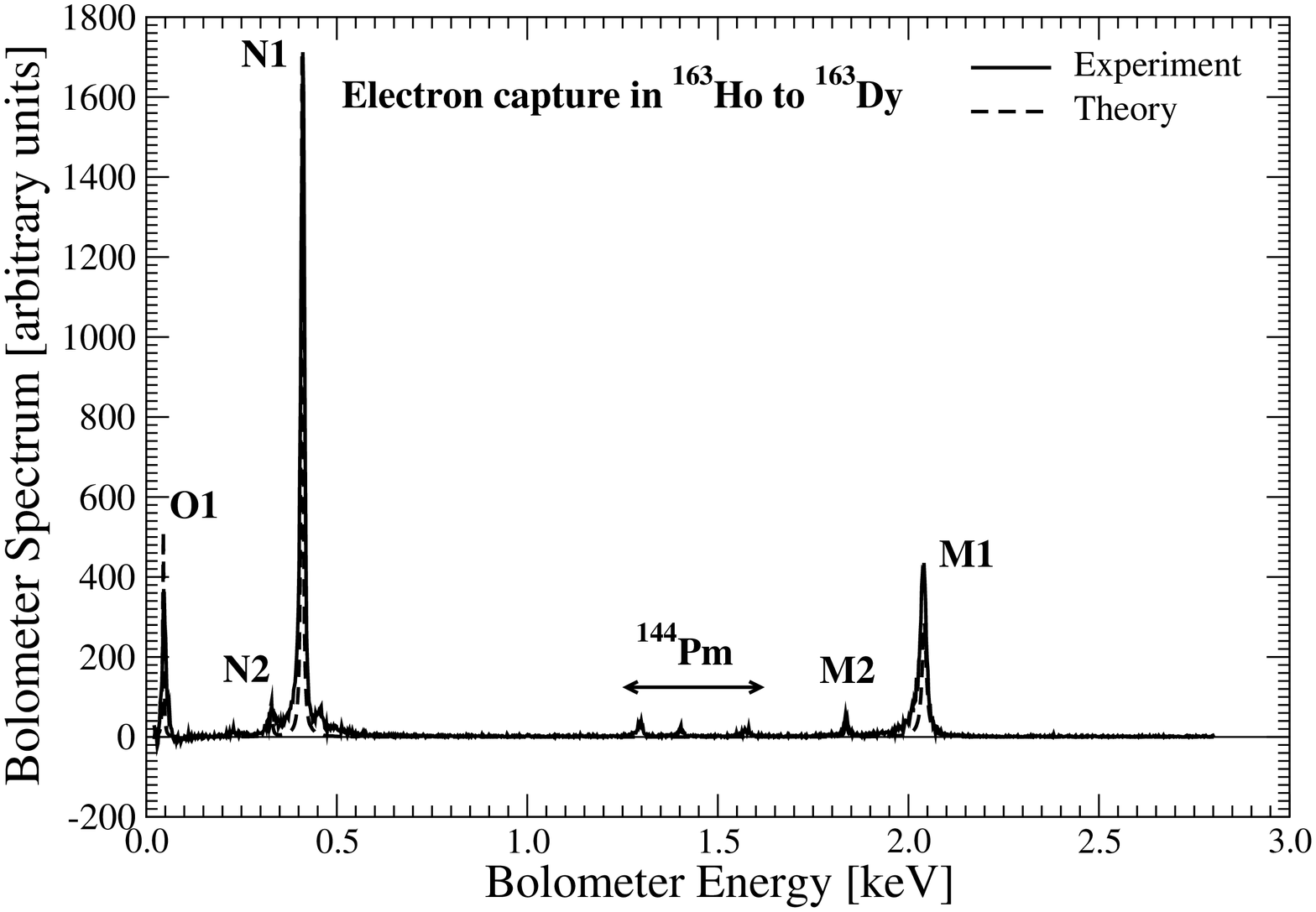,angle=-90,scale=0.6}
\end{minipage}
\begin{minipage}[t]{16.5 cm}
\caption{Experimental (solid line) and theoretical (dashed line) 
linear bolometer spectrum (\ref{decay}) and (\ref{P1}) 
for electron capture in $^{163}Ho$ to $^{163}Dy$ with the  
Q-value = 2.8 keV (see eq. (\ref{QECHo})) including  the one-  
and two-hole probabilities calculated in this work and compared to the 
Heidelberg experiment of Ranitzsch et al.  
\cite{Ra2} and Gastaldo et al.\cite{Lo2} for the total bolometer energy from 0.02 to 2.80 keV. 
The theoretical spectrum is normalized to the data of the N1 (4s1/2) peak at 0.42 keV. 
The three small experimental lines between 1.2 and 1.6 keV 
are originating from small admixtures of $^{144}Pm$. 
The experimental counts are binned in 2 eV intervals. 
Due to background subtraction the number of counts 
in some bins can be negative. In some regions of the energy the number of counts 
per bin are zero or one. Due to negative and zero counts in a bin it is not possible to 
plot the data logarithmically as in figure \ref{FaeRob}, which would show the small effects 
of the two-hole excitations in Dy better. The theoretical spectrum is calculated 
for zero neutrino mass and with the excitation and width of the one hole states 
of the ECHo collaboration \cite{Ra2} and \cite{Lo2}  listed in table \ref{Bind}.  
\label{Lin}}
\end{minipage}
\end{center}
\end{figure}
 
\section{Comparison with Carlson and Nestor}

Thomas A. Carlson, C. W. Nestor, Thomas C. Tucker and F. B. Malik \cite{Carlson1,Carlson2} 
derived by physics arguments for the antisymmetrization and 
the Pauli principle the probability to excite apart of 
a hole in f' also an additional  particle-hole state:

\bea
P_{Carlson; f,p} = ( 1 - (|<(n,\ell,j)_{p',Dy}|(n, \ell,j)_{p,Ho}>|^2)^{N_{n \ell j}} 
-  \hspace{4cm} \nonumber \\ 
\sum_{n'< F; \ne n'_p} \frac{N_{n,\ell,j}\cdot N_{n',\ell,j}}
{2j + 1} 
\cdot |<n', \ell,j (Dy)|n, \ell,j (Ho)>|^2   ) \hspace{3cm}
\label{P5}
\eea

The definition of $N_{n, \ell, j}$ is given in eq. (\ref{P4}). $(n, \ell, j)_{p,Ho}$ and 
$(n, \ell, j)_{p',Dy}$ indicate electrons in the Ho and in the Dy atoms with the same quantum numbers. 

\begin{figure}[tp]
\begin{center}
\begin{minipage}[tl]{18 cm}
\epsfig{file=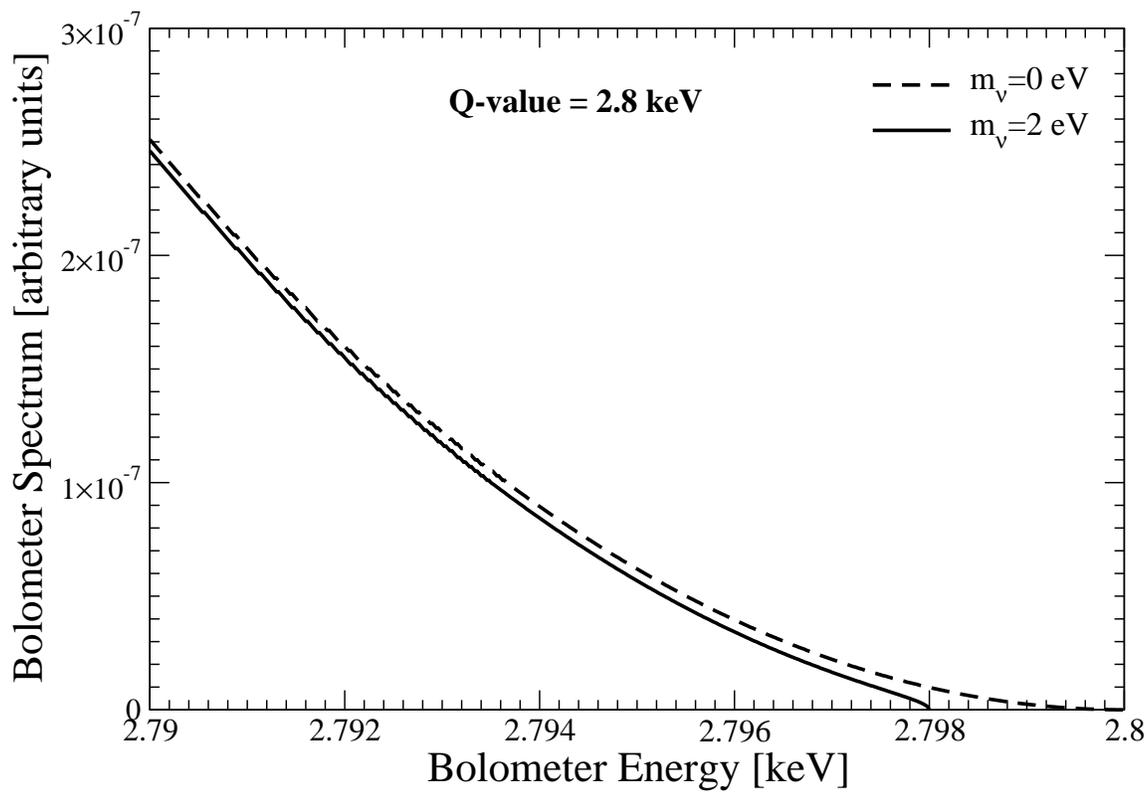,angle=-90,scale=0.6}
\end{minipage}
\begin{minipage}[t]{16.5 cm}
\caption{The upper 10 eV of the bolometer spectrum from 2.790 
to the Q-value = 2.800 keV for neutrino masses 
$ m_{\nu}\  = \ 0.0 \ eV\  (dashed \ line)\  and\  m_{\nu} \ =\  2\  eV\ 
(solid \ line)$. 
\label{Fae0-2}}
\end{minipage}
\end{center}
\end{figure}

\begin{figure}[tp]
\begin{center}
\begin{minipage}[tl]{18 cm}
\epsfig{file=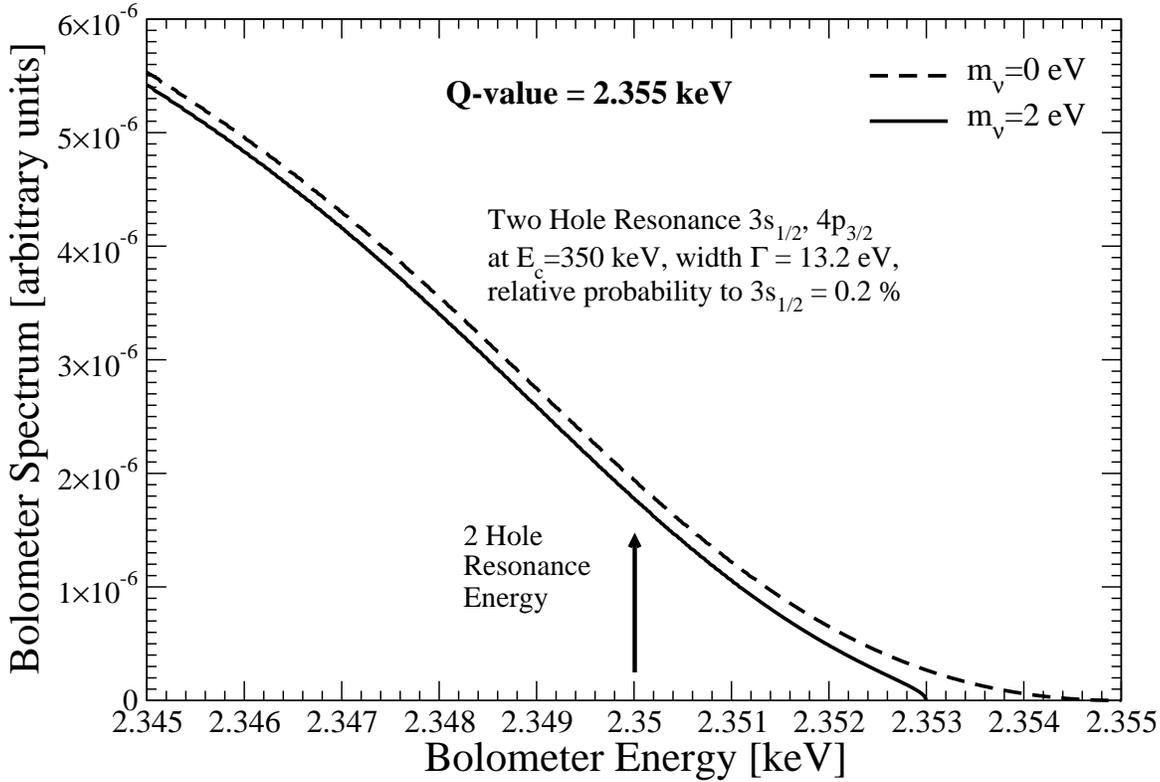,angle=-90,scale=0.6}
\end{minipage}
\begin{minipage}[t]{16.5 cm}
\caption{The upper 10 eV of the bolometer spectrum from 2.345 keV  
to the assumed Q-value = 2.355 keV for neutrino masses 
$ m_{\nu}\  = \ 0.0 \ eV\ (dashed\ line) \  and\  m_{\nu} \ =\  2\  eV\ 
(solid\ line)$. 
The two hole state $3s1/2,\  4p3/2 \ at\  E_c\  =\  2.350\  keV\ $ 
in $^{163} Dy$ is just below the assumed Q-value = 2.355 keV 
within the width $\Gamma \ =\ 13.2\ eV$.  
In a simultaneous fit of the neutrino mass and the Q value also the position,
 the width and the strength of the resonance state must be included. A finite neutrino mass 
produces at the upper end  of the spectrum a special fingerprint, which can not 
be produced  by a resonance state. The fit to the upper end of the spectrum is 
hoped to show this fingerprint as finite neutrino mass. 
\label{ZweiQ2355-0-2}}
\end{minipage}
\end{center}
\end{figure}

\begin{figure}[ht]
\begin{center}
\begin{minipage}[ht]{18 cm}
\epsfig{file=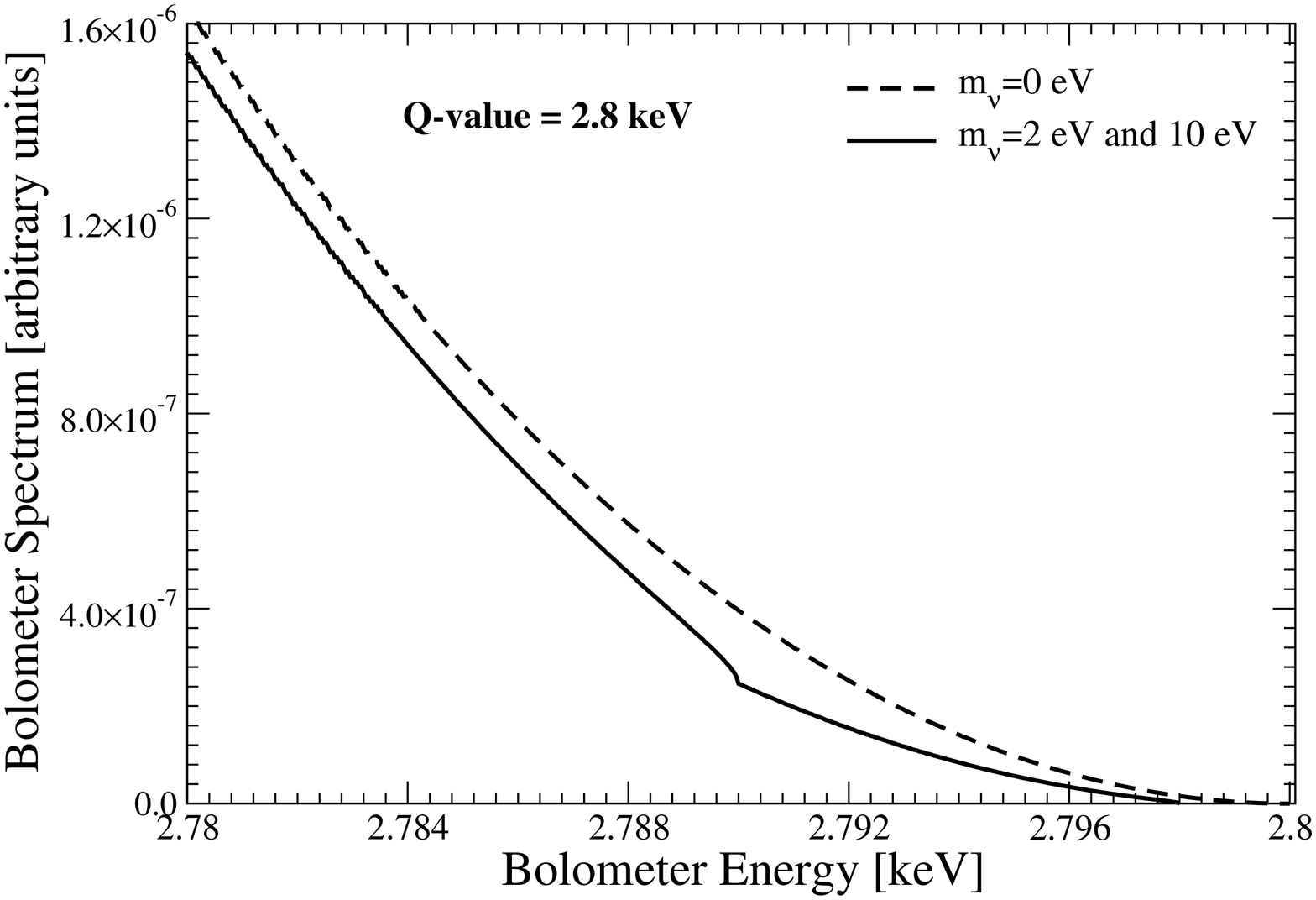,angle=-90,scale=0.6}
\end{minipage}
\begin{minipage}[t]{16.5 cm}
\caption{The upper 20 eV of the bolometer spectrum in $^{163}Dy$ for 
a neutrino with mass $m_ \nu \ = \ 0 \ eV $ (dashed line) and an electron neutrino, 
which is a mixture of two mass eigenstates with   
$ m_{\nu}\  = \ 2 \ eV\  and\  m_{\nu} \ =\  10\  eV$ (solid line) 
for the assumed Q-value = 2.8 keV. The mixing coefficients 
are adopted from \cite{Capozki} assuming only mixing of two neutrinos with probabilities of 
$ U_{e,1}^2(2\  eV) \ =\  0.636 \ and \ U_{e,2}^2(10\ eV) \ =\  0.364 $. 
The neutrino with $m_\nu\  =\  0 \ eV$ is assumed to 
have $U^2_{e,1} \ = \ 1.0 $ and $U^2_{e,2}\  =\  0.0$. 
The onset of the heavy neutrino of $m_\nu \ = \ 10\  eV$ 
at $(Q\  - \ 10)\  eV \ = \ 2.790\  keV$ can be seen in the theoretical spectrum. Finite 
experimental energy resolution will at least partially smear out this effect.   
\label{Zwei2-10}}
\end{minipage}
\end{center}
\end{figure}

The two expressions (\ref{P1}) from this paper and ( \ref{P5}) 
by Carlson and Nestor \cite{Carlson2} are in the 
Vatai  approximation \cite{Vatai1,Vatai2} identical. 
That means one neglects exchange terms and assumes, that  the overlaps of equivalent electron orbitals  
with the same quantum numbers in the parent and in the daughter atom are equal to unity. Thus we have 
in the Vatai approximation \cite{Vatai1,Vatai2}  verified the formula 
of Carlson and Nestor \cite{Carlson2} including here more rigorously the 
antisymmetrization and the Pauli principle and showing also expressions  beyond the Vatai approximation.

The logarithmic (with basis 10) bolometer Spectrum (\ref{decay}) and (\ref{P1}) 
for the one and two hole probabilities of figure \ref{FaeRob} is calculated   
with the parameters of Robertson \cite{Robertson} and a  Q-value = 2.8 keV for
the energy between 0.0 and 2.8 keV. 
In figure \ref{Lin} the one-hole energies and the widths are used from table \ref{Bind}. 
The excitations of one-hole states in
$^{163} Dy$ after electron capture in $^{163} Ho$ and   
also the excitations of two-hole states in Dy are included.   
For the width in this work as far as they are not known experimentally 
the estimates of Robertson \cite{Robertson} are assumed to allow a better comparison. 
In a linear plot of the bolometer spectrum the two hole excitations are hard to see and 
the spectrum calculated in this work and the one calculated with the probabilities of 
Robertson  \cite{Robertson}, Carlson and Nestor \cite{Carlson2} look almost identical, 
although the two hole probabilities are in some cases strongly 
different as shown in tables \ref{TwoHoles} and \ref{TwoHoles2}. The two hole 
modifications of the bolometer spectrum show up clearly in a logarithmic plot, but are 
suppressed in a figure with a linear ordinate.

In figure \ref{Lin} the theoretical results for the bolometer spectrum 
in electron capture in $^{163}Ho$ to $^{163}Dy$ of this work are compared 
with experimental data of Ranitzsch et al. \ \cite{Ra2} and of Gastaldo et al. \ 
\cite{Lo2} in a linear plot assuming a Q value of 2.8 keV.

In figure \ref{Fae0-2} the upper end (2.70 to 2.80 keV) of the theoretical linear bolometer 
spectrum of the present approach including one- and two-hole states are shown for an 
electron neutrino mass of 0.0 eV and 2 eV. A similar result is 
obtained for Q values Q = 2.3 keV and Q = 2.5 keV. 

What happens, if the Q-value falls 
within the width of a two hole resonance? 
This situation is displayed in figure  \ref{ZweiQ2355-0-2}.
The two hole resonance 3s1/2, 4p3/2 lies at 2.350 keV with a width 
$\Gamma \ = \ 13.2 \ eV$. The figure shows the upper part of the bolometer 
spectrum from 2.345 to 2.355 keV for a neutrino mass 
of $m_\nu\  = \ 0\  eV$ and $ m_\nu\  = \ 2\  eV $. 
                                                
Figure \ref{Zwei2-10} shows the upper end of a theoretical 
spectrum with a mixture of two mass eigenstates for the electron 
neutrino. The mixing probabilities are adopted from Capozki et al. \cite{Capozki}. 
The onset of the 10 eV admixture at 2.790 keV can be seen.
\vspace{0.5cm}

The importance of excited states in Dy for the neutrino 
mass determination does not depend, if it is a one-, 
a two- or a multi-hole state or if it is of a different nature. 
Under the assumption, that the shape of 
the resonances are Lorentzian, the importance of a specific state 
for the determination of the neutrino mass can be seen from eq. (\ref{decay}).
A measure for the importance of a state for the determination 
of the neutrino mass is:

\begin{eqnarray}
 \mbox{importance} \propto \frac{B_{f} \cdot \Gamma_{f'}}{(Q - E_{f'})^2 + \Gamma_{f'}^2/4} 
  \approx \frac{B_{f} \cdot \Gamma_{f'}}{(Q - E_{f'})^2 }
\label{importance}
\end{eqnarray}

The dependence  on the resonance energy $E_{f'}$ is 
$(Q - E_{f'})^{-2}$, if the distance of the energy to the 
Q value is larger than the width. So states near to the 
Q value have normally the largest influence on 
the determination of the neutrino mass. In general one 
needs a simultaneous fit of the neutrino mass,
the Q value and the parameters ($E_{f'}$, $B_{f'}$\  and the width $\Gamma_{f'}$) 
of the most important resonance (or even resonances). 
This makes the determination of the neutrino mass very difficult or perhaps even impossible. 

To analyze how many and which parameters must be fitted to the upper end of the experimental spectrum, 
we introduce the definitions $\Delta E_{C}$ and $\Delta E_{f'}$ 
in eq. (\ref{parameters}) and we assume, 
that only one resonance determines the upper end of the spectrum near the Q value 
and that the profile of this line is Lorentzian. 

\begin{eqnarray}
 E_C = Q - \Delta E_C; \hspace{2cm} E_{f'} = Q - \Delta E_{f'}  
\label{parameters}                                              
\end{eqnarray}

$\Delta E_C$ describes the variable energy and $\Delta E_{f'}$ the distance of the leading resonance to the
Q value, and 
$\Gamma_{f'}$ is the width of this resonance. 

\begin{eqnarray} 
 \frac{d\Gamma}{dE_c} \propto (Q - E_c) \cdot\sqrt{(Q-E_c)^2 -m_{\nu}^2} 
\cdot \frac{S}{(E_c - E_{f'})^2 +\Gamma_{f'}^2/4}  \nonumber \\
= \Delta E_C \cdot \sqrt{\Delta E_C^2 - m_\nu^2} \cdot \frac{S}{(\Delta_{f'} - \Delta E_C)^2 + \Gamma_{f'}^2/4}
\label{end}
\end{eqnarray}
        
Here $ S \ \propto \ \lambda_0 \cdot B_f \cdot \Gamma_{f'}$ is the strength of the resonance. 

An estimate shows, that the one hole states play the decisive role for the behavior at the Q  value assuming, that the more accurate value of Q = 2.8 keV of the ECHo collaboration \cite{Ra} is correct. The relative weight 174 for the highest one-hole state 
 3s1/2 at 2.0418 keV with $P_{1-hole}\ = \ 100\ \%$ is at the Q value, the important area for the neutrino mass, by a factor 100  larger than the weight 1.6 of the highest two-hole state at 2.4742 keV with $P_{2-hole}\ = \ 0.167\ \%$ from table \ref{TwoHoles}.

\begin{eqnarray} 
 Relative\  weight  \propto  \frac{P_{1-hole}\ \%}{(Q - E_{f'})^2} 
= \frac{100\ \%}{(2.80 - 2.04)^2} = 174   \nonumber \\
Relative\  weight  \propto  \frac{P_{2-hole}\ \%}{(Q - E_{f'})^2} 
= \frac{0.167\ \%}{(2.80 - 2.47)^2} = 1.6 
\label{weight}
\end{eqnarray}

The widths of the one- and the two-hole states of the highest energies are assumed to be the same $\Gamma \ = \ 13.2\ eV$ (
see table \ref{TwoHoles}) and thus are not changing the relative weights.  
This means, for a Q value of Q = 2.8 keV the 2-hole states seem not to play the dominant role for the determination of the neutrino mass at least judging from the states of energies closest to the Q values.  
                                            
Under the assumption, that one resonance determines the upper end of the spectrum and 
that the shape of this state is Lorentzian, at the Q value assuming, that the one has four parameters to fit simultaneously: 
the neutrino mass $m_\nu$ , the distance of the resonance to the Q value 
  $\Delta E_{f'}$, the strength S and the width
$\Gamma_{f'}$. To include the experimental resolution in the fit, one must first fold the theoretical upper end of the spectrum with the experimentally determined profile of the detector.  
 
\section{Conclusions}
In the present work the bolometer spectrum after electron capture in $^{163}Ho$ for 
the deexcitation of $^{163}Dy$ has been 
calculated including the one- and two-hole excitations in Dy. 
The main improvements compared to Robertson \cite{Robertson} and to Carlson and Nestor 
\cite{Carlson2} are: The two hole probabilities are calculated in the atoms 
Holmium and Dysprosium directly involved in electron capture, by which one wants 
to determine the neutrino mass. Robertson \cite{Robertson} used 
for electron capture in Z=67 Holmium results for Z=54 Xenon 
calculated 
by Carlson and Nestor \cite{Carlson2}. 

The present work takes also into account 
selfconsistently in the relativistic Dirac-Hartree-Fock approach the different 
hole states in $^{163} Dy$. So for each one hole state the remaining 65 electron 
wave functions are calculated selfconsistently and used to determine the two hole probabilities.  
The larger number of electrons in Dy than in Xe allows 
additional two hole states, which previously 
have not been included. The two hole probabilities in Dy calculated here are quite 
different from the probabilities of Robertson \cite{Robertson}, Carlson and Nestor
 \cite{Carlson2} calculated in Z=54 Xenon. To test the numerical results of this work 
three two-hole probabilities have also been calculated by hand.  

The neutrino mass must be 
determined by a simultaneous fit together with the Q value and the properties of the relevant resonances (Assuming a Lorentzian profile these are positions, strengths and widths.) to the upper end of the spectrum.
The finite neutrino mass 
provides at the upper end of the spectrum a characteristic deviation from the  usual line shape, which can not be simulated by 
a resonance in Dy. This fingerprint close to the Q value should show up in the fit to the data. 
The finite experimental energy resolution has to be folded into the theoretical spectrum before one fits it to the data. Thus an excellent  resolution of the measurement near the Q value is essential.  \newline \newline

Acknowledgment: I want to thank the members 
of the ECHo collaboration and especially Loredana Gastaldo for making me available 
the experimental bolometer spectrum of figure \ref{Lin}.

\end{document}